# Silicate clouds and a circumplanetary disk in the YSES-1 exoplanet system

K. K. W. Hoch[1]*, M. Rowland[2], S. Petrus[3,4,5], E. Nasedkin[6,7], C. Ingebretsen[8], J. Kammerer[9], M. Perrin[1], V. D'Orazi[10,2], W. O. Balmer[8], T. Barman[11], M. Bonnefoy[12], G. Chauvin[6], C. Chen[1], R. J. De Rosa[13], J. Girard[1], E. Gonzales[14], M. Kenworthy[15], Q. M. Konopacky[16], B. Macintosh[17,18], S. E. Moran[11,3], C. V. Morley[2], P. Palma-Bifani[19,20], L. Pueyo[1], B. Ren[19], E. Rickman[21], J.-B. Ruffio[16], C. A. Theissen[16], K. Ward-Duong[22], Y. Zhang[23]

[1]Space Telescope Science Institute, 3700 San Martin Dr, Baltimore, MD 21218, USA
[2]The University of Texas at Austin, 110 Inner Campus Dr, Austin, TX 78712, USA
[3]NASA-Goddard Space Flight Center, Greenbelt, MD 20771, USA
[4]Instituto de Estudios Astrofísicos, Facultad de Ingeniería y Ciencias, Uni. Diego Portales, Av. Ejército 441, Santiago, Chile
[5]Millennium Nucleus on Young Exoplanets and their Moons (YEMS), Santiago, Chile
[6]Max-Planck-Institut für Astronomie, Königstuhl 17, 69117 Heidelberg, Germany
[7]Trinity College Dublin, College Green, Dublin 2, Ireland
[8]Johns Hopkins University, 3100 Wyman Park Drive, Baltimore, MD 21218, USA
[9]European Southern Observatory, Karl-Schwarzschild-Straße 2, 85748 Garching bei München, Germany
[10]University of Rome Tor Vergata, Via Cracovia, 50, 00133 Roma RM, Italy
[11]Lunar & Planetary Lab, University of Arizona, 1200 E University Boulevard, Tucson, AZ 85721, USA
[12]University of Grenoble Alpes, 621 Av. Centrale, 38400 Saint-Martin-d'Hères, France
[13]European Southern Observatory, Alonso de Córdova 3107, Vitacura, Región Metropolitana, Chile
[14]San Francisco State University, 1600 Holloway Avenue, San Francisco, CA 94132
[15]Leiden Observatory, Leiden University, Sterrenwachtlaan 11, 2311 GW Leiden, Netherlands
[16]UC San Diego, 9500 Gilman Drive, La Jolla, CA 92093, USA
[17]University of California Observatories, 550 Red Hill Rd, Santa Cruz, CA 95064, USA
[18]University of California Santa Cruz, 1156 High St, Santa Cruz, CA 95064, USA
[19]Observatoire de la Cote d'Azur, 96 Bd de l'Observatoire, 06300 Nice, France
[20]LESIA, Observatoire de Paris, Univ PSL, CNRS, Sorbonne Univ, Univ de Paris France
[21]European Space Agency (ESA), ESA Office, Space Telescope Science Institute, 3700 San Martin Drive, Baltimore, MD 21218, USA
[22]Smith College, 1 Chapin Way, Northampton, MA 01063, USA
[23]California Institute of Technology, 1200 E California Blvd, Pasadena, CA 91125

*Corresponding author(s). E-mail(s): khoch@stsci.edu

**Young exoplanets provide a critical link between understanding planet formation and atmospheric evolution[1]. Direct imaging spectroscopy allows us to infer the properties of young, wide orbit, giant planets with high signal-to-noise. This allows us to compare this young population to exoplanets characterized with transmission spectroscopy, which has indirectly revealed the presence of clouds[2-4], photochemistry[5], and a diversity of atmospheric compositions[6-7]. Direct detections have also been made for brown dwarfs[8-9], but direct studies of young giant planets in the mid-infrared were not possible prior to JWST[10]. With two exoplanets around a solar type star, the YSES-1 system is an ideal laboratory for studying this early phase of exoplanet evolution. We report the first direct observations of silicate clouds in the atmosphere of the exoplanet YSES-1 c through its 9-11 μm absorption feature, and the first circumplanetary disk silicate emission around its sibling planet, YSES-1 b. The clouds of YSES-1 c are composed of either amorphous iron-enriched pyroxene or a combination of amorphous $MgSiO_3$ and $Mg_2SiO_4$, with particle sizes of ≤0.1 μm at 1 millibar of pressure. We attribute the emission from the disk around YSES-1 b to be from submicron olivine dust grains, which may have formed through collisions of planet-forming bodies in the disk.**

The YSES-1 (TYC 8998-760-1, 2MASS J13251211-6456207) system consists of two Jovian planets around a young, solar mass star. Located at 94 parsec in the Sco-Cen star forming region[11-13], these widely separated planets at ~160 au and 320 au projected separation (1.6 and 3.2 arcsec angular distance) exhibit favorable contrast ($3 \times 10^{-3}$ and $1 \times 10^{-4}$ in K-band), providing an ideal laboratory for spectroscopic reconnaissance of a low mass ratio (0.005) exoplanetary system. The estimated masses are 14±3 and 6±1 $M_{Jup}$ based on evolutionary models for the 16.7±1.4 Myr system age[11]. Located near the "L/T" transition, both planets are much redder than other exoplanets and field brown dwarfs[11], suggesting they have distinct atmospheric conditions or processes. At near-infrared (NIR) wavelengths, YSES-1 b is accessible via ground-based spectroscopy, whereas YSES-1 c is too faint for such measurements. K-band (~2.5 μm) spectroscopy of YSES-1 b revealed molecular features from $H_2O$, CO, and the first direct $^{12}CO/^{13}CO$ isotope ratio measurement. Moreover, observations of Hα line emission from YSES-1 b indicate gas accretion onto the planet, implying the on-going formation of this system [14,15].

We present new JWST observations of YSES-1 b and c, consisting of low resolution (R~100) spectra from 0.6 to 12 microns (see Figure 1). Both planets are visible in the raw data without any need for starlight suppression (see Methods for details of observations and data processing). The spectra for both planets show clear signatures of CO, $H_2O$, $CO_2$, and $CH_4$ from 1–5 μm, and $H_2O$ absorption from 5–7 μm. Notably, for the outer planet YSES-1 c, the absorption feature seen from 9–12 μm is consistent with silicate clouds (Figure 1). As a late-L (~L7.5) type object,

YSES-1 c is expected to have silicate clouds, similar to other mid- to late-L type brown dwarfs [16-19], though the shape of its silicate feature is distinct from that of comparable objects (see Figure 2).

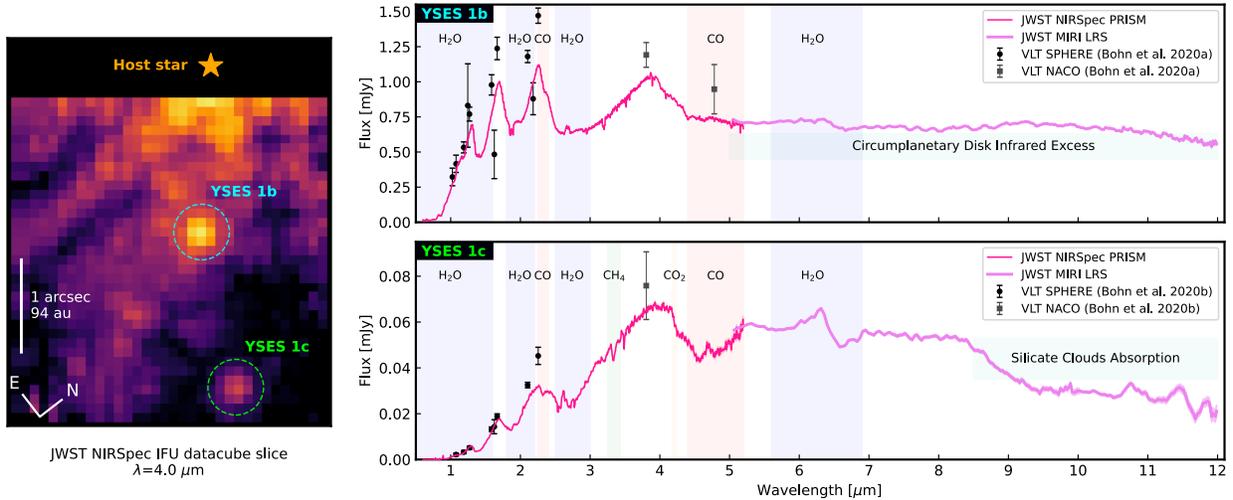

To characterize the physical and atmospheric parameters for each planet we conduct two independent analyses, using ForMoSA[20,21] coupled with the Exo-Rem[25] grid to forward model the spectra and petitRADTRANS (pRT)[23] to perform atmospheric retrievals (Methods). With high SNR and broad wavelength coverage, both methods are able to infer precise measurements of physical parameters. However, the dominant source of uncertainty is the systematics in the models, which are not reflected in the measured posterior distributions[24]. For completeness, we present the statistical uncertainties associated with each model in Extended Data Table 3, though we adopt preferred parameter ranges spanning the measurements from each method in order to reflect the model uncertainty.

Beginning with YSES-1 c, we find that no existing grid of cloudy, self-consistent atmospheric models (e.g. Exo-Rem[22]) reproduces YSES-1 c's silicate feature. Based on the combined forward modelling and retrieval analysis, we find an effective temperature range of 950–1100 K and a surface gravity range of 3.0–3.7 dex. From a composition standpoint, our analyses yield a metallicity ([M/H]) range of 0.27–0.52 and an atmospheric carbon-to-oxygen (C/O) ratio between 0.60–0.72. Up to 18% of oxygen can be sequestered in the silicate clouds in brown dwarfs[25], which assumes 3.28 oxygen atoms per silicon atom condensing in $MgSiO_3$ and $Mg_2SiO_4$[26]. Accounting for this, the resultant bulk C/O ratio from the atmospheric retrievals is 0.65. Overall, the ForMoSA analysis indicates solar composition for YSES-1 c's atmosphere while the pRT retrieval hints at modestly enriched abundances, comparable to other directly imaged exoplanets [27].

Late L-type field brown dwarfs[8] and companions such as VHS 1256 b[9] share similar deep silicate absorption features in the mid-infrared. In comparison, YSES-1 c has a unique silicate absorption feature as shown in panels 1 and 2 in Figure 2. The absorption begins at longer wavelengths compared to those objects, closer to 8.5 μm rather than 7.5-7.8 μm for the brown dwarfs or 8.0 μm for VHS 1256 b. The shape and depth of the feature are also distinct. Using the cloudless atmosphere grid ATMO[28], we measure the silicate equivalent width indices[24] (See Supplemental Methods). We find that YSES-1 c has a large silicate index (panel 3 of Figure 2), following the trend of stronger and redder features for lower-mass and younger objects. The depth of the feature is consistent with YSES-1 c sharing an equator-on inclination with its host (81°±9° [29,30]), where the cloud absorption strength is expected to be maximised[8]. This confirms that silicate clouds in young and low surface gravity objects are different from field-age free floating brown dwarfs and substellar companions, previously hypothesized based on exoplanet colors[31] and silicate indices[8].

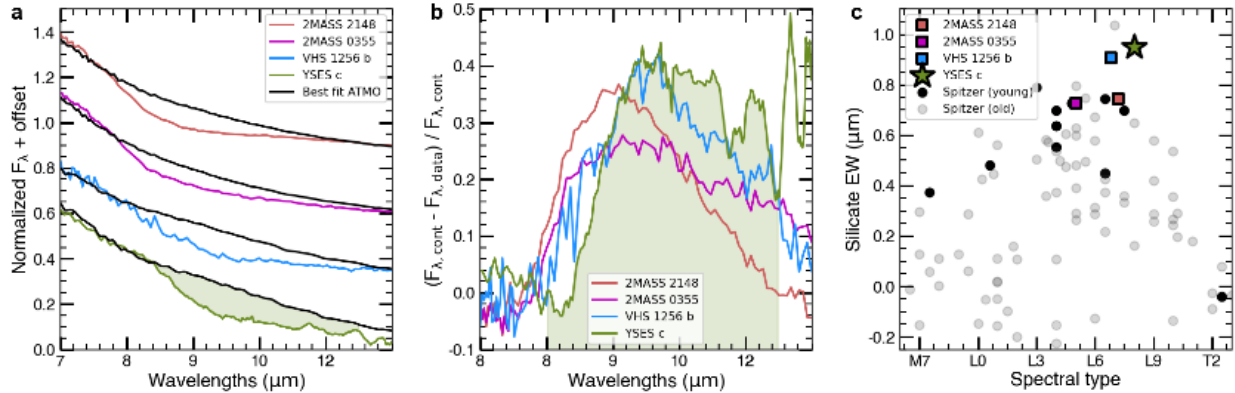

We explore the physical mechanisms underlying this difference by generating a suite of custom cloud models that include a variety of silicate species and fit them to the data longwards of 7 μm. We assume atmospheric properties based on our reported ranges reported and vary the cloud composition, mean grain size of the particles, and cloud height in the atmosphere. We prioritize fitting the first half of the feature from 8.5–10 μm where the SNR is higher than at longer wavelengths. We fit three amorphous $SiO_2$, $MgSiO_3$, and $Mg_2SiO_4$ compositions, as shown in Figure 3. The best fit grain size distribution was tightly centered on 0.1 μm and the best cloud location was at 1 millibar of pressure for all cloud compositions tested. $MgSiO_3$ was the best fit, but a slight wavelength shift remained between the cloud model and the observed feature. We therefore fit additional pyroxene cloud models with differing fractions of iron (Fe) to shift the feature[32]. We also fit models which include two clouds with differing fractions of amorphous $MgSiO_3$ and $Mg_2SiO_4$. Finally, we fit models using the Distribution of Hollow Spheres (DHS) approximation, which redshifted the feature too far. Either a small amount of iron in the cloud particles or a combination consisting of 60-90% amorphous $MgSiO_3$ matches the data better than amorphous $MgSiO_3$. See Figure 3 bottom right, and see Extended Figure 6 in Methods for further details of cloud modeling. These results are consistent with iron and magnesium hydrides condensing into thick clouds with smaller particles being lofted to low pressures[33]. Alternatively to silicate mixtures, it has recently been proposed that polymorphs of silica and other minerals may play a role in shifting cloud spectral features to longer wavelengths[34].

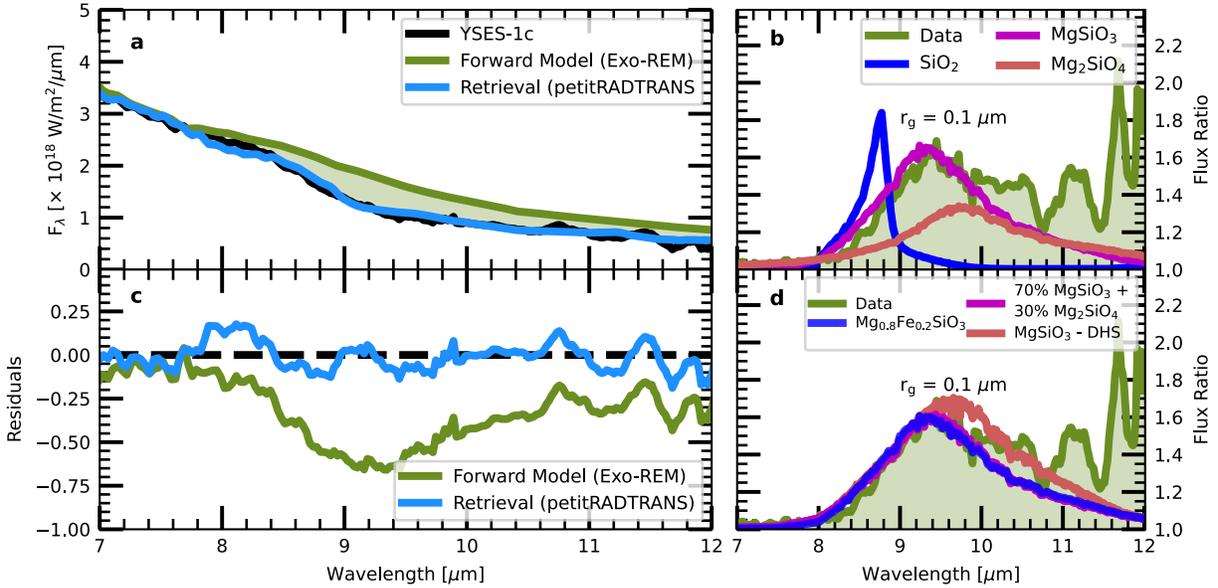

Independently, an atmospheric retrieval also finds that $MgSiO_3$ is the best fit silicate for the clouds in YSES-1 c. We found that crystalline $MgSiO_3$ particles with a mean particle size of 0.20 ± 0.02 μm and a cloud base of 0.02 bar provided the best fit to the spectrum, broadly similar to the detailed modelling properties. An iron-enriched silicate cloud was also tested, with a composition of $Mg_{0.5}Fe_{0.5}SiO_3$, and was found to have a slightly worse $\chi^2/\nu$ (9.86) than the pure enstatite cloud (7.18). For the iron rich composition, the cloud base sinks to 0.4 bar and the particle size increases to 1.7 μm. The difference between the retrieved cloud structure and the detailed model likely stems from degeneracies between particle size and cloud base pressure that current cloud parameterizations cannot yet resolve. Such degeneracies also exist between the cloud parameterisation and the thermal profile[28], and we also find that the temperature profile of YSES 1-c follows a shallow temperature gradient, which also acts to redden the spectrum.

While YSES-1 b's atmosphere is too hot for silicates to condense into clouds, silicate emission from a circumplanetary dust disk (CPD) is visible in the MIRI observations. Previously-detected Brackett-γ emission[14] indicates on-going accretion onto the planet, hinting at the presence of circumplanetary material. Although high-contrast polarimetric observations were conducted to search for circumplanetary dust in this system[35], no disk signal was detected. Our JWST data reveals an infrared excess from 4–14 micron, confirming the presence of hot circumplanetary dust. This places YSES-1 b among the very few substellar companions around which circumplanetary disks are directly observed, including PDS 70 c[36] and GQ Lup B[37].

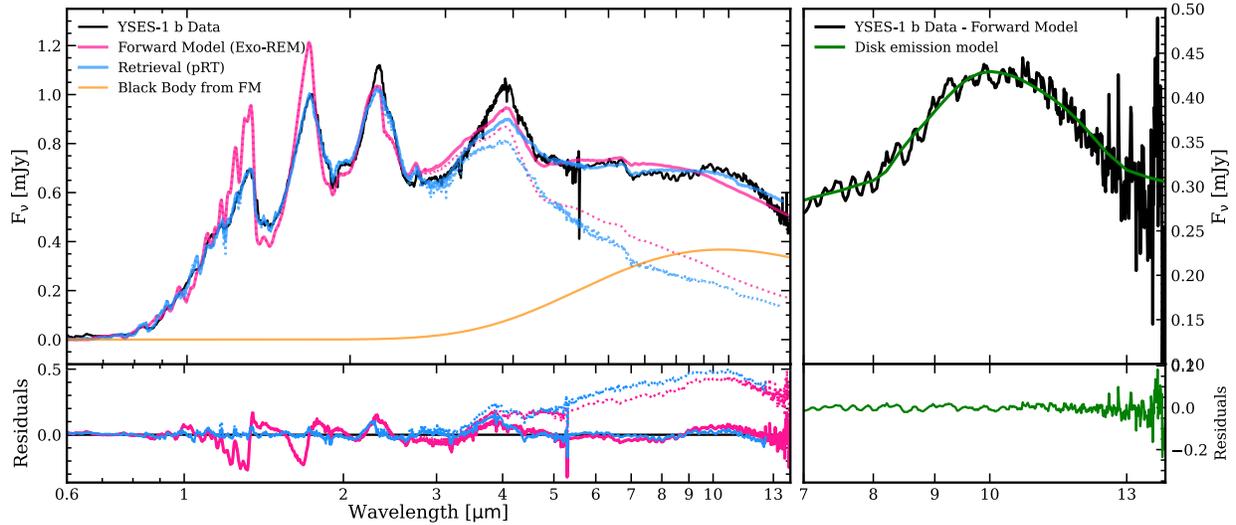

We jointly fit atmospheric models for the planet atmosphere and circumplanetary disk thermal emission. We use the same forward modeling method from above, both with and without an additional blackbody to account for the CPD (see Methods Section "Atmospheric Forward Modeling" and Figure 4). We find an effective temperature range of 1600–2020 K and 2.8–4.8 dex for the surface gravity. The retrievals and forward models do not reproduce the spectra from ~1–2.5 microns, which is particularly sensitive to the surface gravity and metallicity[38]. This discrepancy may be due to the limited understanding of the impact of active accretion of material onto the atmospheres, as this has mostly been studied in protostars[39].

Both our forward models and retrievals find best-fit blackbody temperatures of ~500 K for the CPD, with a disk radius between 8.5–20 $R_{Jup}$. This blackbody model cannot fully replicate the observed emission feature from 8–11 µm. The presence of this broad emission feature indicates the presence of micron-sized silicate grains in the CPD (see panel 2 of Figure 4). This is the first clear detection of silicates in a CPD, in contrast to the blackbody appearance of the small sample of CPDs for which infrared spectra exist [37,40]. In order to quantify the dust properties, we apply physically motivated models which capture the thermal processing of dust grains in the CPD. We isolate the infrared excess by subtracting the planetary atmosphere continuum using the best-fit Exo-REM model. This infrared excess is fit with a disk emission model, assuming the material in the disk emits in thermal equilibrium, has a power-law size distribution resulting from collisional equilibrium, and is composed of olivine ($MgFeSiO_4$). This fit is displayed in the right panel of Figure 4. The emission is consistent with a disk of $MgFeSiO_4$ grains with a minimum size of 0.6±0.04 µm and a temperature of 484.6±7 K for the silicates and $602.1^{+36}_{-56}$ K for the black-body component, corresponding to expected thermal equilibrium for dust 12-35 $R_{Jup}$ away from the planet. The estimated total dust mass for all particles smaller than 1 mm is ~$1.1 \cdot 10^{-9}\ M_{Earth}$ ($8.8 \cdot 10^{-8}\ M_{Moon}$).

No other CPD has been found to have such small, hot grains to date[37]. The absence of small grains could be due to dust grains having already grown to sizes larger than 5 μm, which is seen in young systems[41,42]. For the Solar System's protoplanetary disk, it is hypothesized that silicate grains experience thermal processing over time, for example from colliding planetesimals, resulting in fine grain sizes. Given the age of the YSES-1 system, the presence of small, hot, olivine grains could indicate that we are seeing a later generation of thermally processed grains caused by collisions of larger satellite forming material in the CPD [43,44].

Our analyses have revealed both the complex atmospheric and environmental features of the YSES-1 system, and the challenges in modelling the on-going physical processes. We confirm the predicted presence of silicate clouds at high altitudes in YSES-1 c, establishing that those clouds are responsible for the extreme reddening of its spectrum. Identifying the detailed composition and structure of cloud particles is a critical step in accounting for all oxygen sinks in an atmosphere. Measuring the cloud properties allows us to connect the atmospheric and bulk C/O, thus placing the planets in context with their host star (See Supplemental Methods). YSES-1 b's circumplanetary material could be second generation grains resulting from collisions of larger grains or the formation of moons, but a deeper understanding of the physics behind the emerging populations of CPDs is required. Future studies of silicate clouds and circumplanetary material will continue to rely on the high SNR and high resolution spectra only obtainable via direct imaging spectroscopy, but a deeper understanding of systematics in atmospheric models is necessary to fully exploit such datasets.

# Figure Legends

**Figure 1: Observed spectra of YSES-1 b and c.** Left panel: A single wavelength (4 μm) slice from the NIRSpec IFU prism datacube, illustrating the geometry of the system. Both planets can be directly seen as point sources within the square instrumental field of view, along with a point spread function halo of glare from the host star which was located just outside of the IFU entrance aperture as indicated. This PSF halo was modeled and subtracted in data analyses prior to spectral extraction for the planets. Right panels: The observed spectra of both planets. Solid

lines show the spectra measured with JWST NIRSpec and MIRI; shaded regions on either side of the line indicate the 1-sigma measurements uncertainties, but these are generally small, comparable to the thickness of the line. Symbols indicate prior ground-based photometric measurements from the literature, also with 1 sigma uncertainties. The colored bands label the major molecular and continuum components.

**Figure 2: Semi-empirical analysis of the silicate absorption feature.** The silicate feature of YSES-1 c is compared to the features seen in three examples of young substellar objects: VHS 1256 b, 2MASS 2148, and 2MASS 0355, ordered by decreasing mass. Left: The observed spectra of each object (colored lines) along with models for cloud-free atmospheres from the ATMO model grid (black lines). Note the shift towards longer wavelengths for the lower mass objects, particularly for YSES 1c. Center: The fractional reduction of emission at each wavelength due to the silicate clouds, computed as the flux ratio between the absorption depth and the cloud-free continuum model. The shaded region highlights the wavelength range used to calculate the silicate index. For clarity, this is shown only for YSES-1 c. However, the same range has been applied to all MIR spectra from JWST and the Spitzer library to ensure consistency in the analysis. Right: Comparison of the silicate equivalent width indices of these objects and a broader sample of brown dwarfs from the Spitzer library. The two young substellar companions, YSES 1c and VHS 1256 b, have among the highest silicate indices compared to the field sample.

**Figure 3: YSES-1 c spectral comparison against silicate cloud models.** The spectrum of YSES-1 c centered on the silicate feature with the best fit cloudy Exo-REM spectrum and best fit retrieved spectrum [upper left], along with their residuals [lower left]. The flux ratio between the best fit Exo-REM model and the data [shaded green] compared to the flux ratios between cloud free models and cloud free models with silicate clouds of different composition are shown in the right two panels. All cloud models shown have a mean cloud particle radius of 0.1 μm and cloud base location of 1 millibar of pressure.

**Figure 4: YSES-1 b spectral comparison against forward model, retrieval, and thermal disk emission model.** The spectrum of YSES-1 b is plotted in black against the best fit cloudy Exo-REM spectrum with an added blackbody and without an added blackbody in pink, the best fit petitRADTRANS retrieval with and without an added blackbody in blue, and the best fit blackbody curve in orange on the left panel with residuals between the fits below. The best fit thermal emission model is shown in green against the Exo-REM model without the added blackbody subtracted from the data divided by the error in the data in black.

# Methods
**Observations, Data Reduction, and PSF Model Subtraction**

The data presented are from the Cycle 1 observing program 2044 (PI: Hoch), with details of instrument settings and exposure times given in Extended Data Table 1. Observations were obtained using NIRSpec IFU Prism mode. We placed the IFU field of view to include both planets with the bright host star outside. However, 2 of the 4 dither positions placed YSES-1 c on the edge of the field of view. For MIRI, the Low Resolution Spectroscopy mode observed the two planets sequentially using a two-point nod.

We performed initial reduction steps using the JWST data pipeline version 1.13.4 and CRDS context 1125. For the NIRSpec data, we performed pixel outlier detection using all 4 dithered exposures, then produced the final spectral datacube using only the 2 dithers where YSES-1 c was within the field of view (see Extended Data Figure 1). To remove the host star point spread function (PSF), we conducted PSF subtraction using models generated with WebbPSF[38]. We fit the 3.0 micron slice of the datacube and iterated to optimize via least squares the alignment between model and host star location. We then iterated over all wavelengths to generate model PSFs and subtract them from the science data. Each wavelength was fit with a multiplicative scale factor to minimize the residuals over a region around each planet to produce a PSF-subtracted datacube (Extended Data Figure 1).

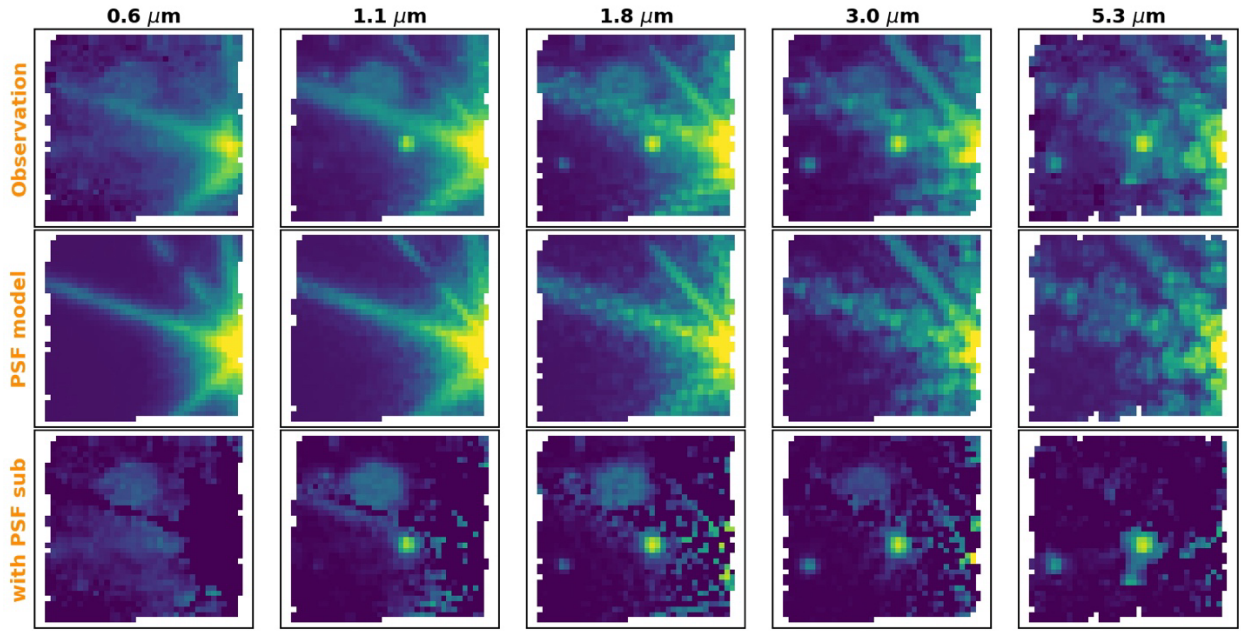

For the MIRI LRS data, the PSF subtraction relied on removing a scaled numerical model of the PSF. We developed a custom code to forward model MIRI LRS data, available on Github at https://github.com/mperrin/miri_lrs_fm. We fit a forward model of the planet as a point source within the slit, the wings of the offset host star outside of the slit, and the diffuse thermal sky background. We iterated to optimize position offsets and flux scale factors. We generated a series of monochromatic PSFs, shifted at each wavelength according to the spectral dispersion profile, and summed to generate a synthetic 2D spectrum. We scaled the flux by a model of the host star's spectral energy distribution and subtracted the model from the science data, and fit an overall flux scale factor and background offset varying with wavelength. The two nods were subtracted from one another to remove the observatory thermal background. The PSF subtraction resulted in a significant decrease in systematics between the two nods (See Extended Data Figures 2-3).

| Target(s) | Instrument and mode | Date | Dithers/Nods | N. Groups | Total Exposure Time (s) | Description |
| --- | --- | --- | --- | --- | --- | --- |
| YSES-1 b & c | NIRSpec Prism IFU | 6/19/2023 | 4 | 17 | 1050* | Both companions in FOV |
| YSES-1 b | MIRI LRS | 4/25/2023 | 2 | 90 | 500 | YSES-1 b in slit |
| YSES-1 c | MIRI LRS | 4/25/2023 | 2 | 300 | 5000 | YSES-1 c in slit |

**Spectral Extraction**

We extracted the spectra from the PSF-subtracted NIRSpec datacube by PSF-fitting photometry using a 2D gaussian PSF model plus a constant background term. We fit the position of each planet after summing over the wavelength axis. Next, we iterated over wavelength and fitted a 2D gaussian + constant at each wavelength while holding the position fixed. Bad pixels were masked and not included. We summed the 2D gaussian model flux at each wavelength to obtain companion spectra. We drew 1000 values from the Gaussian model parameter posterior, its uncertainties from the least squares fit. We evaluated the companion flux for each of these 1000 draws and reported lower and upper flux uncertainties based on the 16th and 84th percentiles and propagated the JWST data reduction pipeline uncertainties.

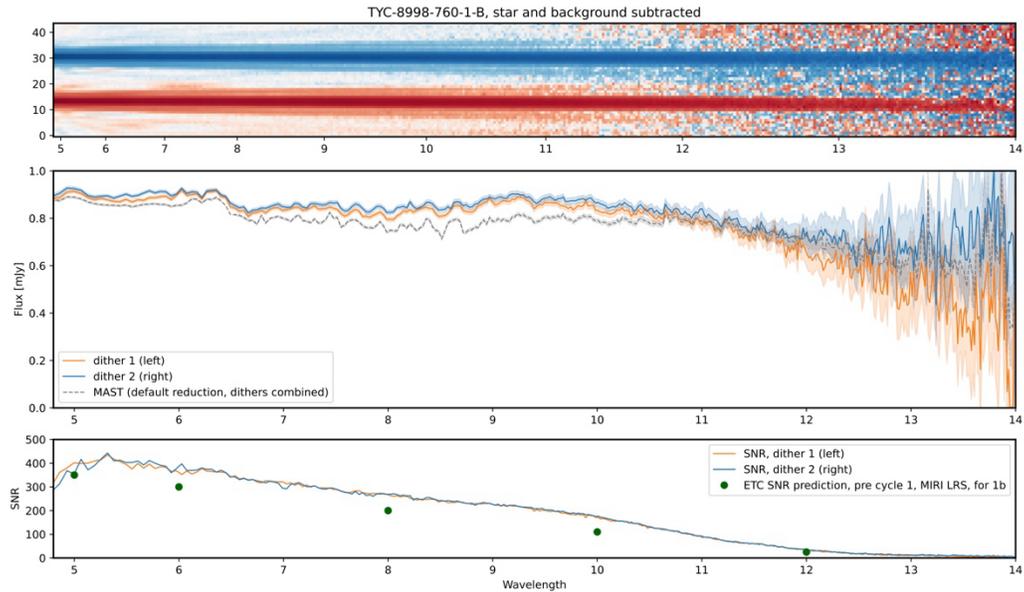

For the MIRI data, we extracted spectra using PSF-profile-weighted extraction[39]. We again generated a dispersed PSF model of the companion in each nod and the model PSF cross-dispersion spatial profile at each wavelength was used to compute a weighted sum of the 2d spectral image. The resulting spectra for each nod were then averaged and uncertainties propagated. This was repeated for both planets. Note that the sensitivity of MIRI LRS decreases sharply beyond 12 μm resulting in lower S/N at longer wavelengths.

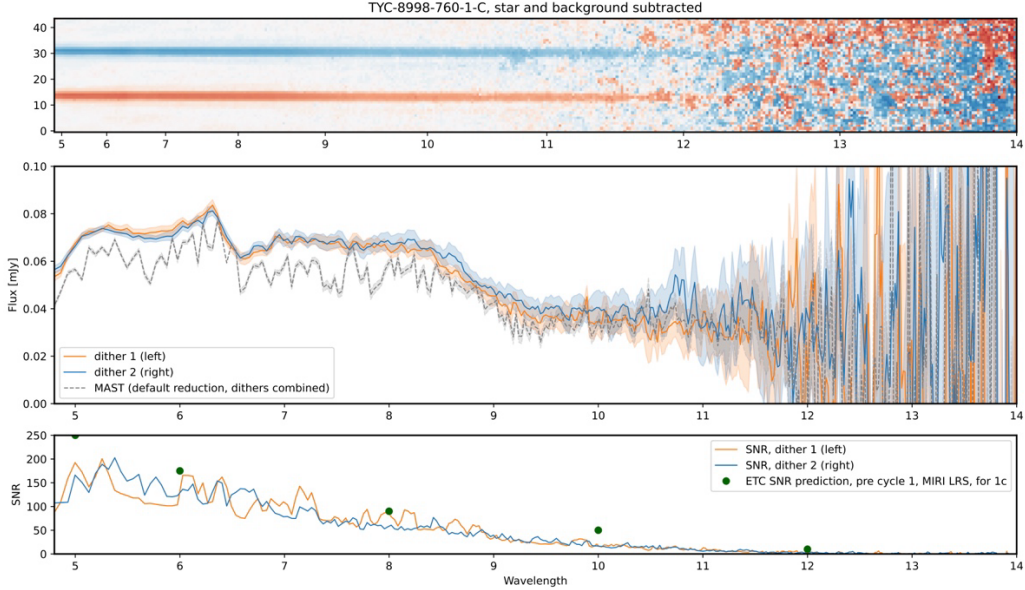

**Atmospheric Forward Modeling**

We forward model the atmospheric spectra using the self-consistent cloudy atmospheric model grid Exo-REM[25]. Exo-REM assumes radiative-convective equilibrium, non-equilibrium chemistry for a limited number of molecules (CO, $CH_4$, $CO_2$, and $NH_3$), and the formation of clouds (iron, $Na_2S$, KCl, silicates, and water). Exo-REM covers $T_{eff}$ = 400 to 2000 K and includes molecular absorptions with the rovibrational bands of $H_2O$, $CH_4$, CO, $CO_2$, $NH_3$, $PH_3$, TiO, VO, and FeH. The grid also includes atomic absorptions with resonant lines from Na and K, and collision-induced absorptions of $H_2$–$H_2$ and $H_2$–He. The particle size distributions are log-normal with a constant effective variance of 0.3. We explore ranges of surface gravities (log(g) = 3.0 to 5.0 dex), metallicities ([M/H] = -0.5 to 0.5) and carbon-oxygen ratios (C/O = 0.1 to 0.8).

We used the ForMoSA code[23,24] with a nested sampling algorithm (pyMultinest[41]) and optimized grid interpolations. The resolution and spectral coverage of the grid are adapted to the data through a Gaussian convolution followed by binning. The luminosity dilution factor $C_k = \left(\frac{R}{d}\right)^2$ is calculated for the synthetic spectra considering d = 94.6±0.3 pc[42]. We combine the atmospheric model with a circumplanetary disk (CPD) model for YSES-1 b consisting of a blackbody defined by a temperature ($T_{CPD}$) and a radius ($R_{CPD}$). All priors used are in Extended Data Table 2.

The entire spectral range was used for YSES-1 b and wavelengths less than 7.5 μm were used for YSES-1 c. However, at medium resolution, self-consistent models struggled to reproduce the data over a wide spectral range inducing under-estimated errors. These errors reflect the good signal-to-noise of the data propagated through the Bayesian inversion, but the errors are dominated by systematic errors in the self-consistent models. To mitigate this, we progressively increasing the uncertainties until the reduced $\chi^2$ of the previous fit reached 1. If the reduced $\chi^2$ fell within the range of 0.5 to 1.5, we adopted the corresponding error values for the data and the posterior distributions.

| Parameter | Prior |
|---|---|
| ExoREM FM | |
| $T_{eff}$ [K] | U(400, 2000) |
| log(g) [dex] | U(3.0, 5.0) |
| [M/H] | U(-0.5, 0.5) |
| C/O | U(0.1, 0.8) |
| $T_{disk}$ [K] | U(1, 1000) |
| $R_{disk}$ [$R_J$] | U(0, 100) |
| pRT Free Retrievals | |
| Mass [$M_J$] | U(0.5,12) |
| Radius [$R_J$] | U(0.7,3.0) |
| $T_{bottom}$ [K] | U(2000,12000) |
| $dlnP/dlnT_1$ | N(0.25,0.025) |
| $dlnP/dlnT_2$ | N(0.25,0.045) |
| $dlnP/dlnT_3$ | N(0.26,0.05) |
| $dlnP/dlnT_4$ | N(0.20,0.05) |
| $dlnP/dlnT_5$ | N(0.12,0.045) |
| $dlnP/dlnT_6$ | N(0.07,0.07) |
| $dlnP/dlnT_7$ | N(0.0,0.1) |
| $dlnP/dlnT_8$ | N(0.0,0.1) |
| $dlnP/dlnT_9$ | N(0.0,0.1) |
| $dlnP/dlnT_{10}$ | N(0.0,0.1) |
| $\sigma_{LN}$ | U(1.05,3.0) |
| $\log(K_{zz})$ [log cm$^2$ s$^{-1}$] | U(3.0,13.0) |
| $f_{SED,X}$ | U(0.0,10.0) |
| $\log(r_{cloud})$ [cm] | U(-7.0,10.0) |
| $f_{cloud}$ | U(0,1) |
| $\log(P_{cloud})$ [bar] | U(-6,3) |
| $\log(X_i)$ | U(-12,-0.3) |
| $b$ | U(-40,-20) |
| pRT CPD | |
| $T_{disk}$ [K] | U(10,1000) |
| $R_{disk}$ [$R_J$] | U(1,100) |

**Atmospheric Retrievals**

We used petitRADTRANS (pRT) version 3.2.0a16 to perform atmospheric retrievals on the planet spectra[26,43]. pRT relies on the pyMultinest [41,44,45] implementation of Nested Sampling[46] to sample the parameter space, estimate the posterior probability distributions and calculate the model evidence. We used 1000 live points with a sampling efficiency of 0.3 for model comparison of $10^7$–$10^8$ models. All priors used are in Extended Data Table 2. We used the entire wavelength range for YSES-1 b, and removed a low SNR region from 0.6-0.88μm for YSES-1 c.

Our model consists of a temperature profile, a chemical model, and a cloud parameterisation, at 134 discrete pressure levels spaced between 1000 bar and $10^{-6}$ bar, with a higher resolution grid at the location of cloud condensation[40,47]. The temperature profile used is taken from Zhang et al. 2023[48], where the free parameters of the model are the temperature gradient at 10 equidistant points in pressure space[48]. Our parameterisation extended the retrieved gradients to the top of the atmosphere, rather than isothermal at pressures lower than $10^{-3}$ bar.

We used a free chemistry approach where the molecular mass-fraction abundances are free parameters constant with altitude. The remaining atmosphere is a mixture of 76% $H_2$ and 23% He. For YSES-1 c we included $H_2O$[49], CO[50], $CH_4$[51], $CO_2$ [52], $NH_3$[53], HCN[54], $H_2S$[55], $PH_3$[56], FeH[57], Na[58], and K[59]. TiO[60] and VO[61] are included for the higher-temperature YSES-1 b. We include collisionally induced absorption from $H_2$–$H_2$ and $H_2$–He, and Rayleigh scattering from $H_2$ and He. The chemical abundances measured resulted in metallicity relative to solar[62] using

$$[M/H] = \log_{10} \frac{\sum_i N_i}{\sum_i N_{i,\odot}},$$

For number fractions $N_i$ and summing over all elements present in the molecular species. The C/O ratio was calculated using the total number of retrieved carbon and oxygen atoms. We exclude the alkali metals, as the retrieved Na abundance is unphysically large due to the steep NIR slope possibly impacted by cloud absorption.

We retrieve the mean cloud particle size, the width of log-normal distribution of particle sizes, and the mass fraction at the cloud base decreasing as a power law with a slope of $f_{SED}$. We included both an iron cloud and a silicate cloud. We compared Mie-scattering, crystalline $MgSiO_3$ [64], and $Mg_{0.5}Fe_{0.5}SiO_3$ [64] compositions to determine the best fit mid-infrared absorption feature. An amorphous $MgSiO_3$ composition was also tested, performing worse than the crystalline particles. We retrieve a patchiness parameter for the silicate cloud, while the deeper iron cloud is assumed to completely envelope the planet[65].

Lastly, we retrieve parameters of mass and radius to calculate the surface gravity and compute the luminosity dilution factor with d=94.6±0.3 pc[42]. For YSES-1 b, we include the same circumplanetary disk model from the forward model fits.

The emission spectrum is calculated using correlated-k opacities binned to a spectral resolution of 400 then convolved to instrumental spectral resolving power. The resolving power is defined for each wavelength channel, and subsequently binned to the instrumental wavelength grid [66]. We only present the most favored models in this work.

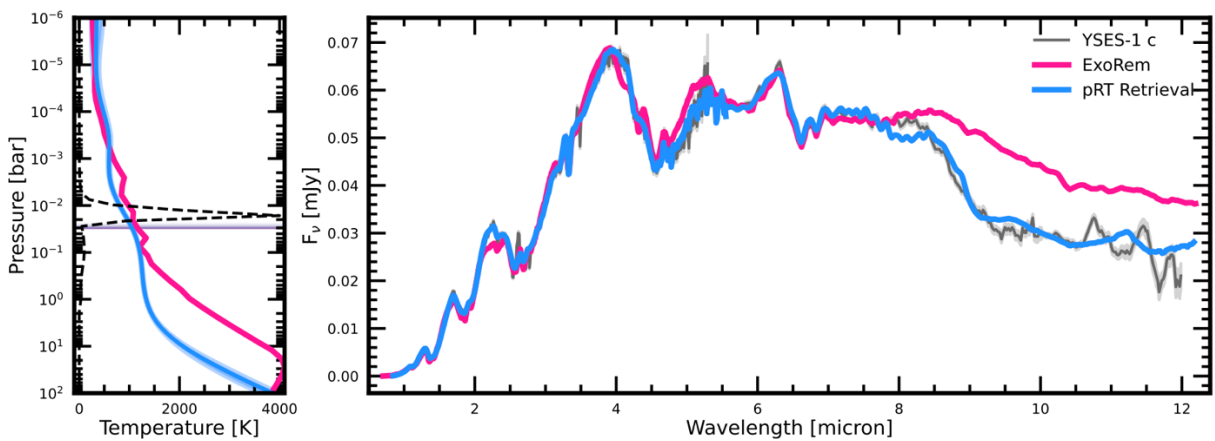

Without error inflation, we measure a lower mass for YSES-1 c (1.80±0.07 $M_J$), driven by the low surface gravity (3.41±0.02 dex). In Extended Data Figure 4, we show the retrieved temperature profile and emission spectrum of YSES-1 c. The effective temperature measured was 1025±1 K and the log bolometric luminosity was -4.7376 ±-0.0003 dex relative to solar. We find a modest enrichment of metals ([M/H] = 0.52±0.01), and a super-solar gas-phase C/O ratio of 0.798±0.003. Roughly 20% of oxygen can be sequestered in silicate clouds [29,30,68]. Accounting for this sink, we find a bulk C/O ratio of 0.64. We additionally find [C/H]=0.66±0.01, [O/H]=0.50±0.01, and [S/H]=0.81±0.01. Carbon and oxygen are moderately enriched compared to the solar value, while the sulfur abundance ratio is strongly enriched. We measure abundances for each of $H_2O$, CO, $CH_4$, $CO_2$, $H_2S$ and K, with the mass fraction abundances (Extended Data Figure 5).

For YSES-1 b, the retrievals were unable to constrain physically plausible values due to the interactions of the planet and the disk. Therefore, we only report the effective temperature ($2002^{+13}_{-9}$ K), the log bolometric luminosity (-3.2612±-0.0007 dex) and the disk temperature (515 ± 4 K) and radius (17.5±0.2 $R_J$) using a blackbody.

The uncertainties on these measurements reflect the precision of the JWST data, and the reduced $\chi^2$ of our best fit for YSES-1 c is 7.18. In order to account for model uncertainty, we performed retrievals with error inflation, where $\sigma_{total} = \sqrt{\sigma_{measured}^2 + 10^b}$, where $b$ is a retrieved parameter [67]. For NIRSpec, we find $b_{NIRSpec}$=-36.823±0.013, where the typical order of the measured uncertainties is 1x10$^{-19}$ W/m$^2$/micron. The inflation term is more dominant for MIRI data, $b_{MIRI}$=-36.765±0.021, with the typical scale of the measured MIRI uncertainties of ~2x10$^{-20}$

W/m²/micron. Error inflation causes lower precision on the measured parameters ($\chi^2/\nu$ near 1). The bulk metallicity decreases to approximately solar and the planet mass decreases to an unphysically small estimate of 0.588±0.02 $M_J$. This method allows a better goodness-of-fit, but the parameters retrieved present unphysical results. Models accounting for 3D effects, improved line lists, and more detailed cloud parameterizations will be required to fully exploit JWST data.

| Model | $T_{eff}$ [K] | log $g$ [cm s$^{-2}$] | [M/H] | C/O | $R_{pl}$ [$R_J$] | $T_{disk}$ [K] | $R_{disk}$ [$R_J$] | Wavelength range [μm] | nParam | DoF | $\chi^2/\nu$ |
|---|---|---|---|---|---|---|---|---|---|---|---|
| **YSES-1 c** | | | | | | | | | | | |
| FM with ExoREM only | 958-10+20 | 3.63-0.06+0.07 | 0.27-0.08+0.11 | 0.61-0.01+0.01 | 1.48-0.04+0.05 | N/A | N/A | < 7.5 | 7 | N/A | N/A |
| FM with ExoREM + Blackbody | 984-17+29 | 3.69-0.13+0.19 | 0.34-0.10+0.08 | 0.61-0.05+0.03 | 1.41-0.05+0.05 | 49-35+38 | 37-24+32 | 0.6-14.0 | 9 | N/A | N/A |
| pRT Retrieval | 1025 ± 1 | 3.41 ± 0.02 | 0.52 ± 0.01 | 0.798 ±0.003 | 1.351 ± 0.003 | N/A | N/A | 0.88-12.0 | 35 | 1086 | 7.18 |
| Inflated Retrieval | 1095.0± 1.5 | 3.05± 0.02 | 0.06± 0.03 | 0.723 ±0.008 | 1.161 ± 0.003 | N/A | N/A | 0.88-12.0 | 37 | 1088 | 0.93 |
| Adopted range | 950–1100 | 3.0–3.7 | 0.030–0.38 | 0.60–0.72 | 1.2–1.5 | N/A | N/A | N/A | N/A | N/A | N/A |
| **YSES-1b** | | | | | | | | | | | |
| FM with ExoREM only | 1513-25+36 | 4.63-0.24+0.23 | -0.01-0.17+0.19 | 0.59-0.10+0..15 | 3.38-0.15+0.12 | N/A | N/A | 0.6-14.0 | 7 | N/A | N/A |
| FM with ExoREM + Blackbody | 1607-22+24 | 4.56-0.41+0.26 | 0.20-0.20+0.15 | 0.70-0.08+0.05 | 2.97-0.09+0.08 | 371-53+46 | 7.35-1.74+2.75 | 0.6-14.0 | 9 | N/A | N/A |
| Inflated Retrieval | 2002-9+13 | 2.9-0.1+0.2 | N/A | N/A | 1.82±0.02 | 515.0 ± 4.0 | 17.5±0.2 | 0.6-12.0 | 40 | 1051 | 1.03 |
| Adopted range | 1600–2020 | 2.8–4.8 | 0.00–0.35 | 0.62–0.75 | 1.8–3.1 | 320–520 | 5.6–18 | N/A | N/A | N/A | N/A |
| Thermal IR | N/A | N/A | N/A | N/A | N/A | 602.1-56+36 | 10.5 ± 1.7 | > 7.5 | 5 | 320 | 0.014 |

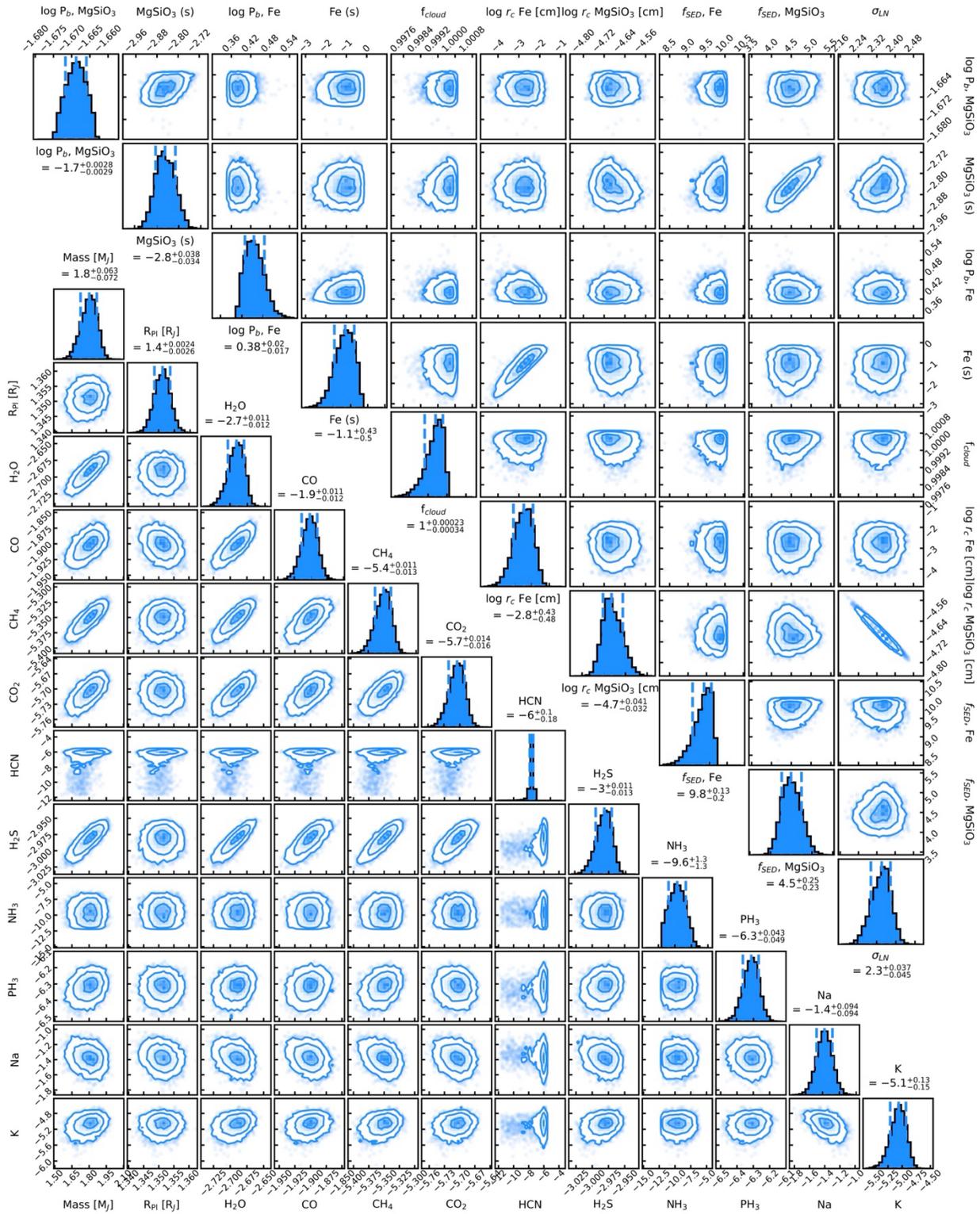

**Silicate Feature Cloud Modeling**

We defined YSES-1 c's silicate feature as a flux ratio between the best fitting cloudy Exo-REM model excluding the silicate feature ($T_{eff}$ = 950 K , log(g)=3.5, [M/H]=0, C/O = 0.60) and the YSES-1 c spectrum.

The silicate feature fitting was performed using VIRGA[69,70] and PICASO[71,72] to model different types of silicate clouds with varying cloud properties. The thermal and chemical profiles from the Exo-REM model were input into PICASO to produce a cloud-free spectrum between 7 and 12 μm. Rather than using the standard EddySed formulation, VIRGA was used to add ad hoc clouds with varying chemical composition, crystalline and amorphous optical properties, particle sizes, cloud base pressures, and cloud column number densities. These clouds with updated optical properties were added to the cloud-free Exo-REM thermal and chemical profiles and PICASO was used to produce new spectra[73]. Silicate features for each cloud model were calculated as a flux ratio between the Exo-REM spectrum with the cloud and the cloud-free Exo-REM spectrum between 7 and 12 μm. YSES-1 c's silicate feature was then compared to all of the silicate features modeled with VIRGA and PICASO.

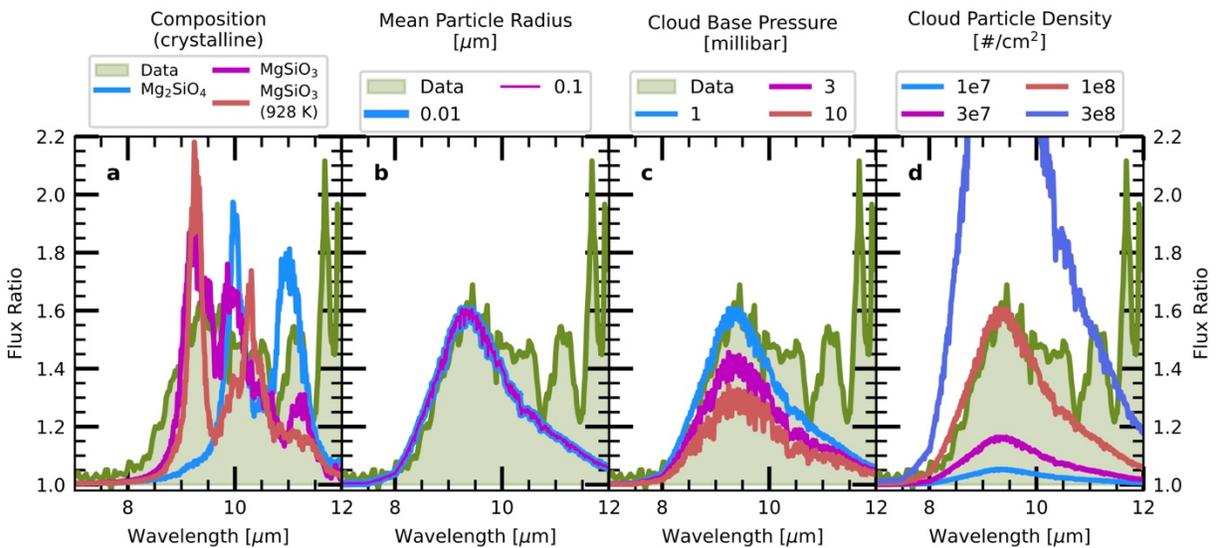

The considered cloud species were $SiO_2$ amorphous[74,75], $Mg_2SiO_4$ (amorphous[76] and crystalline[77]), $MgSiO_3$ (amorphous[78] and crystalline[64]), crystalline $MgSiO_3$ at 3 temperatures (10 K, 300 K, and 928 K)[79], and several amorphous pyroxene species ($Mg_{0.95}Fe_{0.05}SiO_3$, $Mg_{0.8}Fe_{0.2}SiO_3$, $Mg_{0.5}Fe_{0.5}SiO_3$, and $Mg_{0.4}Fe_{0.6}SiO_3$)[78]. Cloud base pressures ranged from 0.1 bar to 0.0001 bar, and the vertical extent of the cloud was set using an $f_{SED}$ value of 1, which was held constant for all models. Cloud column number densities ranged from $10^7$ to $10^9$ particles/cm². Particle size distributions were tightly focused (a log normal distribution with σ = 1.2) around values ranging from 0.01 to 10 μm. Particle size distributions were held constant with pressure. Cloud combination models used a mixture of amorphous $MgSiO_3$ and $Mg_2SiO_4$ with the same cloud base pressures and mean particle sizes, but differing fractions of the total column number density. The combination models fit the observed feature using a mean particle size of 0.1μm, a cloud base pressure of 1 (−0.2, +0.6) millibar, and a column density of 1 (±0.1) × $10^8$ 1/cm². The precise combination of amorphous $MgSiO_3$ and $Mg_2SiO_4$ varied between 60–90% $MgSiO_3$ based on the other parameters. The mass density of cloud particles at the base of the cloud for the best fit cloud models quoted above were between 2.1–3.7 × $10^{-7}$ g/cm³. We note that mass density at the base of the cloud can also be achieved with smaller particle sizes and larger column densities, so the mean particle size of 0.1 µm is only an upper limit. Additional models with the particle size distribution determined by a Hansen distribution did not change the color of the feature or the inferred composition. Extended Data Figure 6 shows the impact of changing each of these parameters on the resulting silicate feature. Clouds with mean particle radius of 1 µm and 10 µm blocked almost all flux in these wavelengths and resulted in flux ratios an order of magnitude larger than the 0.1 µm and 0.01 µm models and therefore are not included in Extended Data Figure 6. We note that the LRS spectrum signal to noise drops closer to 12 µm due to the faintness of YSES-1 c (See Extended Figure 2). We therefore, only considered the first half of the feature. Only deep MIRI MRS observations can obtain the signal to noise needed to characterize the entire silicate cloud feature beyond 12 µm.

**Empirical Fitting of the Infrared Excess for YSES-1 b**

We fit the infrared excess using a size distribution of silicate grains from sub-micron to millimeter-sized. We subtracted an Exo-REM model containing only the atmosphere emission to isolate the IR excess. We omitted data at wavelengths < 7 µm as the water vapor in the atmosphere creates absorption at 6.3 µm. There is a clear 10 µm emission feature consistent with micron-sized silicate grains, explaining why simple blackbody emission did not fit the excess[80]. Blackbody grains are large (>10 µm) and their thermal emission produces a broad feature. Solid-state emission features are more narrow and indicate the presence of smaller sub-micron to micron-sized dust. We constrain the dust grain size distribution by fitting the emission excess assuming that the grains are spherical (Mie Theory) and composed of amorphous olivine ($MgFeSiO_4$[78]) as it is the most common silicate species detected in the ISM and protoplanetary disks[81]. While crystalline silicates are observed in T Tauri disks, these silicates have distinctive sharp spectral features not observed in our spectrum[82,83].

We model the thermal emission assuming that the dust is composed of both small silicate grains and large black body grains. The particle size distribution for the small grains is a power law consistent with collisional equilibrium using:

$$F_\nu = C_1 \int_{a_{min}}^{a_{max}} \pi a^2 a^{-3.5} Q_{abs}(a) B_\nu(T_{sil}) da + C_2 B_\nu(T_{bb}),$$

[98] where a is the grain size, $a_{min}$ is the minimum grain size, $a_{max}$ is the maximum grain size, $Q_{abs}$ is the absorption efficiency, $T_{sil}$ is the temperature of the small silicate grains and $T_{bb}$ is the temperature of the large black body grains. In our fit, we fix $a_{max}$ to 1 mm and use an MCMC to estimate the best fitting values for amin, $T_{sil}$, $T_{bb}$, $C_1$, and $C_2$. We estimate $a_{min}$ = 0.71 µm, $T_{sil}$ = 488 K, $T_{bb}$ = 602 K, $C_1$ = 801, and $C_2$ = 7080.

From the temperature of the silicates, we calculated the distances of grains from the planet by assuming the grains are in thermal equilibrium and heated by YSES-1 b. The distance of the micron-sized grains in the limit that $2\pi a \ll \lambda$, is:

$$d = \frac{1}{2}\left(\frac{T_{YSES-1b}}{T_{sil}}\right)^{\frac{5}{2}} R_{YSES-1b},$$

where $T_{YSES-1b}$ is the temperature of YSES-1 b and $R_{YSES-1b}$ is it's radius. For the blackbody emitting grains:

$d = \frac{1}{2}\left(\frac{T_{YSES-1b}}{T_{sil}}\right)^{2} R_{YSES-1b}.$

The distances of the silicates are $10.5 \pm 1.7\ R_{Jup}$ and the black body grains are $17.6 \pm 3.7\ R_{Jup}$.

The mass of the circumplanetary disk is dominated by the material in large, meter to kilometer-sized objects, but they do not contribute to the 10 μm emission. This feature is caused by small particles < 10 microns. Thus the dust mass calculated for the CPD is from small dust particles, which is a small fraction of the true mass.

To calculate the mass of the circumplanetary disk, the size distribution was integrated:

$M_{dust} = C_1 d^2{}_{YSES-1} \int_{a_{min}}^{a_{max}} \left(\frac{4\pi}{3} a^3\right) a^{-3.5} \rho\, da.$

$C_1$ is the constant of proportionality fit for the micron-sized silicate emission, $d_{YSES-1}$ is the distance to YSES-1b from the observer, $a$ is the particle size, $a_{min}$ and $a_{max}$ are the minimum and maximum grain sizes fit to silicate emission respectively, and $\rho = 3.71\ g/cm^3$ is the mass density of olivine. The resulting dust mass is ~$6.472 \pm 0.4e15$ grams which is $8.8e-8$ times the mass of Earth's moon.

**Code Availability**

This publication made use of the following code software to analyze these data: NumPy[85], astropy[86], matplotlib[87], SciPy[88], pandas[89], ForMoSA[20,21], VIRGA[69,70], PICASO[71,72], pyMultinest[41], WebbPSF[38], petitRADTRANS[23].

**Data Availability**

The data used in this paper are associated with *JWST* program GO 2044 and are available from the Mikulski Archive for Space Telescopes (https://mast.stsci.edu). The data used for host star measurements are associated with UVES/VLT Program: 106.20ZM.00 and XShooter/VLT Program: 103.2008.001 and are available from the ESO Archive (https://archive.eso.org/).

**Methods References**

[38] Perrin, M. D., Sivaramakrishnan, A., Lajoie, C.-P., Elliott, E., Pueyo, L., et al. Updated point spread function simulations for JWST with WebbPSF. Space Telescopes and Instrumentation 2014: Optical, Infrared, and Millimeter Wave, 9143, 91433X (2014)

[39] Horne, K., An optimal extraction algorithm for CCD spectroscopy. Publications of the Astronomical Society of the Pacific, 98, 609-617 (1986)

**Acknowledgements**
S.P. is supported by the ANID FONDECYT Postdoctoral program No. 3240145 and an appointment to the NASA Postdoctoral Program at the NASA–Goddard Space Flight Center, administered by Oak Ridge Associated Universities under contract with NASA. V.D. acknowledges the financial contribution from PRIN MUR 2022 (code 2022YP5ACE)
funded by the European Union—NextGeneration EU. This work is based on observations made with the NASA/ESA/CSA *JWST*. The data were obtained from the Mikulski Archive for Space Telescopes at the Space Telescope Science Institute, which is operated by the Association of Universities for Research in Astronomy, Inc., under NASA contract NAS 5-03127 for *JWST*. These observations are associated with program *JWST*-GO-02044. Support for program *JWST*-GO-02044 was provided by NASA through a grant from the Space Telescope Science Institute.

**Author contributions**
All authors have played significant roles, with some specific contributions as follows. K. H, M. P., Q. M. K, C. A. T., J.-B. R., C. M., E. G., K. W.-D., E. R., L. P., M. B., V. D., S. P., T. B., J. G., R. J. D. R., B. R. and G. C., aided in the development of the original proposal and made significant contributions to the overall design of the program. M. P. and K. H. generated the observing plan with input from the team. M. P. conducted the data reduction and starlight subtraction and performed the MIRI spectral extraction, and J. K. and K. H. co-led the spectral extraction of NIRSpec Prism. M. R. led the cloud modeling effort, S. P. led the forward modeling effort and silicate index empirical analysis, E. N. led the retrieval effort, and C. I. led the thermal modeling with guidance from C. C. P. P.-B. implemented the ability to fit a CPD in our forward modeling framework. M. K., Y. Z., S. E. M., W. B., B. R., R. J. D. R., and B. M. aided in the interpretation and made contributions to the writing of this paper. K. H., M.P., M. R., S. P., E.N., and C. I. generated figures for this paper.

**Competing interests**
The authors declare no competing interests.

**Additional information**

**Supplementary information**
Supplementary information is included with this manuscript as a Supplementary Methods Section.

**Correspondence and requests for materials**
Should be addressed to K. K. W. Hoch.

**Author Information**
*Corresponding author email: khoch@stsci.edu

# Extended Data Figure Legends and Table Titles and Footnotes

**Extended Data Table 1: Overview of JWST Observations.**
*: only half of the total NIRSpec exposure time was used for analyses as noted in the text.

**Extended Data Figure 1: PSF subtraction of NIRSpec IFU Prism data to remove host star light.** The top row shows slices of the combined NIRSpec Prism data cube at 5 different wavelengths. The data orientation shown here is rotated 90º relative to Fig 1. The middle row shows slices of the PSF model of the host star. The bottom row shows slices of the resultant PSF-subtracted data cube, showing clearer detections of the companions without contamination from the host star. The diffuse roughly circular illumination seen in the third row at wavelengths <= 3 microns is believed to be an optical ghost from reflection within NIRSpec; this is not subtracted by the PSF modeling but due to its location it has no impact on the extracted spectra of the two planets.

**Extended Data Figure 2: PSF subtraction and spectral extraction of MIRI LRS data of YSES-1 b.** The top panel shows the two separate nods in blue and red for YSES-1 b illustrating the spectral traces after PSF and background subtraction. The middle panel shows spectral extractions from both traces with 3 sigma errors plotted. The black spectra is the average of the two nods. The dashed spectrum shows the MAST reduced and extracted spectra to demonstrate the systematics removed by our PSF subtraction. The bottom panel shows the SNR over wavelength of the respective extracted spectra from the middle panel, as well as the ETC calculations from the Cycle 1 proposal.

**Extended Data Figure 3: PSF subtraction and spectral extraction of MIRI LRS data of YSES-1 c.** See Extended Data Figure 2 caption for description

**Extended Data Table 2: Priors for ExoREM Forward Model fitting and petitRADTRANS retrieval fitting.**
*U(a,b) is a uniform prior in the interval a,b, while N(μ,σ) is a Gaussian prior with mean μ and standard deviation σ. For the pRT retrieval, the parameters are as described in [50].

**Extended Data Figure 4: Forward model and retrieved spectrum compared to YSES-1 c spectrum.** Left panel: the pressure-temperature profile for the nearest ExoRem grid point to the best-fit parameters, and a 90% confidence region for the pressure-temperature profile as inferred by the pRT retrieval. The dashed line indicates the emission contribution function averaged across wavelength. Most of the flux is emitted between 0.01 and 0.03 bar, just above the location of the silicate cloud layer whose optical depth is indicated by the purple shading. Also shown are representative condensation curves for $MgSiO_3$, $Mg_2SiO_4$ and Fe, all of which are expected to condense deeper in the atmosphere than what is found by the retrieval. Right panel: the best-fit ExoRem forward model and the maximum-likelihood model from the pRT retrieval are compared to the observed spectrum of YSES-1 c.

**Extended Data Figure 5:** Posterior parameter distributions for YSES-1 c as inferred from the pRT retrieval. Not shown are the parameters for the PT profile, which is shown in Extended Data Figure 4. The units of the chemical abundances are in log mass fraction.

**Extended Data Table 3: Summary of forward modeling, retrieval fitting, and thermal modeling.**

**Extended Data Figure 6: Cloud composition, mean particle radius, cloud base pressure, and cloud particle density fits.** Shown in panel 1 are different silicate species of crystalline $Mg_2SiO_4$, crystalline $MgSiO_3$ (averaged over all temperatures), and crystalline $MgSiO_3$ at 928 K; panel 2 shows different particle radii fits; panel 3 shows different cloud base pressures; and panel 4 shows different particle densities, all against YSES-1 c.